# Stability and photometric accuracy of CMOS image sensors in space: Radiation damage, surface charge and quantum confinement in silicon detectors

Michael E. Hoenk

*Jet Propulsion Laboratory, California Institute of Technology, Pasadena, CA 91109[1]*

Stability and photometric accuracy of silicon imaging detectors are essential for the Habitable Worlds Observatory and a range of NASA missions that will explore time domain astrophysics and astronomy over a spectral range spanning soft X-rays through the ultraviolet (UV), visible, and near infrared. Detector stability is one of the oldest and most challenging problems in NASA missions. The challenges are particularly acute in the extreme ultraviolet range, where near-surface absorption of high-energy photons causes surfaces to degrade rapidly. The susceptibility of back-illuminated silicon detectors to ionizing radiation damage is dramatically demonstrated by the Extreme Ultraviolet Imaging Telescope (EIT) currently flying on the joint ESA-NASA Solar and Heliospheric Observatory (SOHO). Soon after launch, the Tektronix TK512CB CCD on EIT suffered severe degradation of charge collection efficiency caused by exposure to solar EUV radiation, resulting in (non)flat-field images with burned-in images of the sun. This had major consequences for the EIT consortium, which needed five years to develop a usable calibration method for the EUV-damaged detector.[1] In the quarter century after EIT's experience with calibrating radiation-damaged CCDs, considerable effort has gone into improving the stability and radiation-hardness of ion-implanted CMOS and CCD imaging arrays.[2] Despite significant improvements to the process, recent observations of quantum efficiency hysteresis (QEH) in Teledyne e2v (Te2v) CCDs raise important questions about the stability of back-illuminated silicon detectors.[3] In this paper, the effects of radiation-induced variability of surface charge on detector stability and photometric accuracy are analyzed in order to assess the implications for future NASA missions.

Before proceeding, we note that consultation with Teledyne-e2v revealed that the ion-implanted detectors tested in Heymes *et al.* are not representative of current device capabilities, and more recent devices are expected to have improved stability. Calculations using the model developed here show that increasing the surface dopant density will improve detector stability. Further study is needed to validate these results with more representative devices. JPL and Teledyne e2v are collaborating on the development and qualification of high-performance UV detectors for spaceflight.[4,5,6]

QE data reported by Heymes *et al.* are reproduced in Figures 1 and 2, together with calculated QE from the model described below. Figure 1 shows the measured QE of a UV-enhanced CCD over the EUV-UV spectral range. Figure 2 shows the QE of the same device measured before and after prolonged exposure to 200nm photons. Data in the figures are compared with a QE model that I developed for this study as a generalization and expansion of the model used in our previous paper.[6] To accommodate arbitrary surface dopant profiles, the detector is divided into *N* regions and the boundary value problem is solved numerically using matrix calculations. Degenerate doping is addressed using bandgap narrowing data from Swirhun *et al.*,[7] which blunts strongly-peaked surface dopant profiles by reducing the potential barrier height and electric field strength near the surface. Finally, calculations of surface recombination velocity in terms of interface and oxide trap densities are based on Shockley-Read-Hall (SRH) statistics applied to semiconductor surfaces, using formulae derived in Andrew Grove's 1967 book on semiconductor physics,[8] and further developed and refined in models of solar cell performance to include an integration over the silicon bandgap.[9] To accomplish this, the model incorporates measurements of cross sections and densities of states *vs* energy for Pb$_0$ traps at the Si-SiO$_2$ interface.[10] Poisson's equation is solved self-consistently to calculate the surface potential as a function of the densities of interface traps ($N_{it}$) and oxide charge ($N_{ot}$). Grove's introduction of an effective surface recombination velocity ($S_{eff}$) is useful as a heuristic explanation of QE instabilities caused by variable oxide charge in radiation-damaged detectors (see Figures 3 and 4). For this study, I've used the more exact formulae for SRH surface recombination in order to investigate the two main sources of surface charge in detectors, interface and oxide traps ($N_{it}$ and $N_{ot}$), which are conflated in models relying on $S_{eff}$. Radiation-induced variability in the occupation of oxide traps is essential for the interpretation of QEH data in Heymes *et al.* and for the following analysis of radiation damage and detector stability in space.

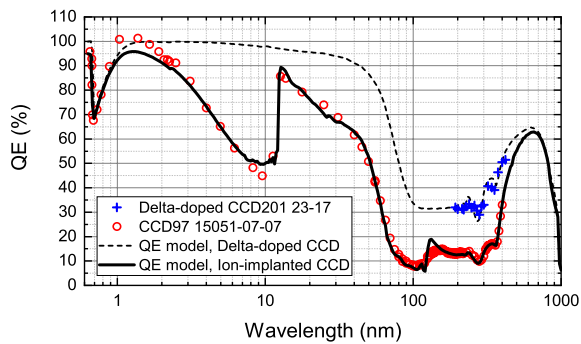

**Figure 1**: Quantum efficiency data for a UV-enhanced CCD97 detector are compared with calculated QE of an ion-implanted detector using trap densities $N_{it} = 3.45 \times 10^{12}$ cm$^{-2}$ and $N_{ot} = 10^{12}$ cm$^{-2}$ as fitting parameters. The QE of a delta-doped CCD is shown for comparison.[6]

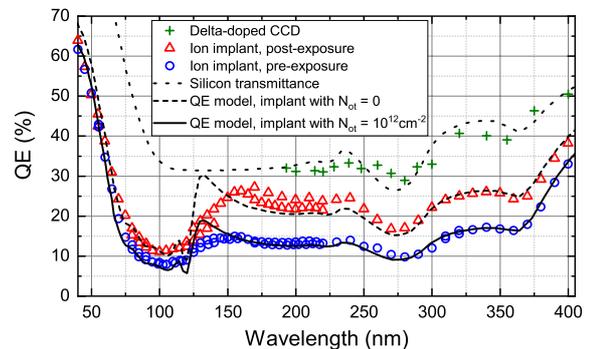

**Figure 2**: QE measurements of the CCD in Figure 1 before and after prolonged UV illumination show a persistent increase in QE caused by exposure to 200nm light.[3] The model shows the measured changes in response are consistent with UV-induced neutralization of positive charge trapped in the oxide.

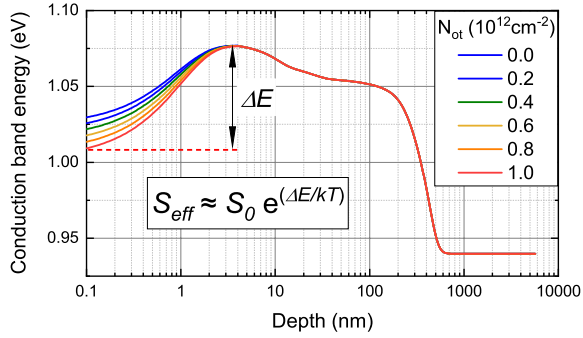
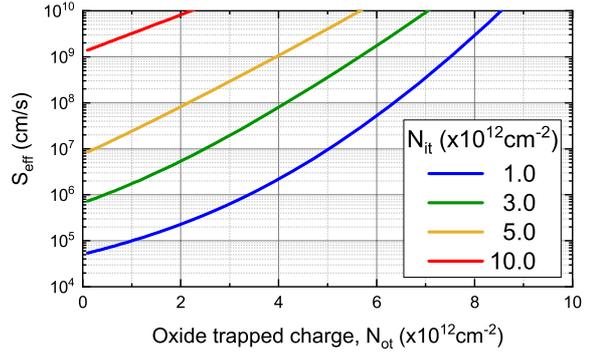

**Figure 3:** The backside potential well of the detector in Figure 1 ($N_{it} = 3.45 \times 10^{12}$ cm$^{-2}$) deepens as the density of charge trapped in the oxide ($N_{ot}$) varies from 0 to $10^{12}$ cm$^{-2}$. Because the *effective* surface recombination velocity ($S_{eff}$) varies exponentially with well depth, small variations in oxide charge can have a large effect on detector QE.

**Figure 4:** Exposure to ionizing radiation causes the Si-SiO$_2$ interface trap density ($N_{it}$) to increase over time. At a given $N_{it}$, the effective surface recombination velocity ($S_{eff}$) varies roughly exponentially with oxide charge ($N_{ot}$). For comparison, see plots of internal quantum efficiency *vs* oxide trapped charge in Figures 5 and 6.

The data in Figure 2 are characteristic of QEH instability, which Jim Janesick described in 1989 as having "plagued the back-illuminated CCD since its invention."[11] The discovery of QEH in state-of-the-art ion-implanted CCDs presents problems and challenges that are important for time domain astronomy. Strategies for the mitigation of QEH instabilities involve flooding the detector with light to charge the detector surface and thereby stabilize the response. In 2013, European Southern Observatory (ESO) astronomers reported that Janesick's UV flood process could be used in ground-based telescopes to improve the UV QE of ion-implanted detectors by up to 50%.[12] In 2010, observations of QEH in Wide Field Camera 3 (WFC3) CCDs motivated the development of a "pinning exposure" that was performed periodically on orbit to neutralize a 4% QE deficit observed after each annealing cycle.[13]

Despite their similarities, the UV flood processes used by ESO astronomers, WFC3, and Heymes *et al.* employ different surface charging mechanisms. Janesick's UV flood process charges the surface while the detector is warm by catalyzing the chemisorption of negatively charged O$_2$ ions on the oxide surface, whereas Heymes *et al.* charged the surface while the detector was cold and under vacuum. In the absence of oxygen, what is causing surface charge to change in Heymes' UV-flood experiment? The answer to this question can be found in a classic experiment performed at Caltech by Carver Mead in 1967, as reported by Snow, Grove and Fitzgerald, which demonstrates UV-induced neutralization of radiation-induced oxide space charge with a threshold photon energy of 4.0 to 5.0 eV.[14]

Based on data and models of radiation-induced degradation of Si-SiO$_2$ interfaces in MOS devices, I propose that the QEH measured by Heymes *et al.* was caused by UV-induced charge injection, which saturates when positive charge trapped in the oxide is neutralized (see Figure 2). In experimental studies of trap-generation dynamics in MOS structures, Nissan-Cohen *et al.* proposed a dynamic charging model based on the idea that oxide charge reaches a steady-state trapping level that depends on the electric field in the oxide.[15] Saturation of QEH in experiments performed by Heymes *et al.* can thus be understood in terms of an equilibrium charge density formed in the oxide.[3] Experimental studies of vacuum ultraviolet (VUV) induced radiation damage in MOS oxides reported by Afanas'ev et al. showed that whereas charge injection into thermal SiO$_2$ is initially slow because of the small cross section of traps in high-quality thermal oxides, ionizing radiation causes accelerated rates of charging and degradation due to "positive feedback in the generation of oxygen vacancies and the clustering of defects, which appear to take place in the degeneration of the MOS system upon VUV irradiation."[16]

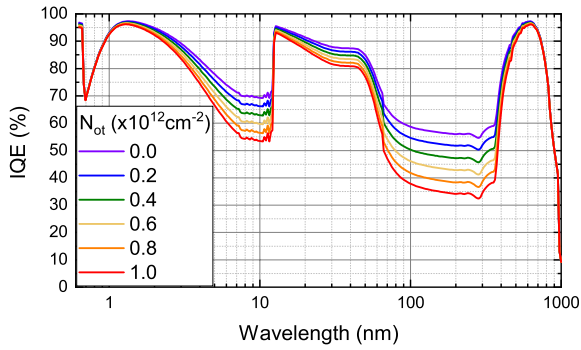 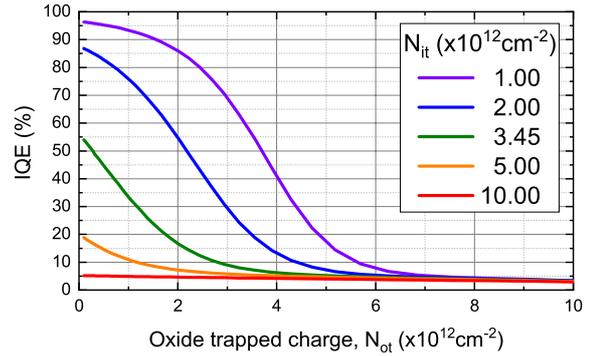

**Figure 5:** The internal quantum efficiency (IQE) of the detector in Figures 1 and 2 ($N_{it} \sim 3.45 \times 10^{12} cm^{-2}$) is very sensitive to variability of charge trapped in the oxide in time and space ($N_{ot}$). For comparison, the quantum efficiency hysteresis (QEH) measured by Heymes et al. corresponds to a change in $N_{ot}$ of approximately $10^{12}$ cm$^{-2}$.

**Figure 6:** Radiation-induced degradation of IQE at 285nm is depicted here in terms of interface and oxide trapped charge. At the beginning of life, the UV QE is high because of low $N_{it}$. Exposure to ionizing radiation damages the oxide, leading to degradation of QE and increased susceptibility QEH (UV-induced variability of oxide charge).

Data and models describing radiation-induced charge injection in MOS oxides, together with calculations using the QE model described in this paper, suggest a causal relationship between radiation damage, oxide charge, and quantum efficiency hysteresis in ion-implanted CCDs. Figure 5 analyzes the spectral response of detectors with different oxide charge densities, using the dopant profile and interface trap density derived for the Heymes et al. CCD (Figure 1). As expected, the greatest variability in internal quantum efficiency (IQE) occurs in the ultraviolet where absorption takes place near the surface, but significant changes are seen across the entire spectral range measured by Heymes et al. Figure 6 extends this analysis by calculating QEH at a specific wavelength (285nm) as a function of interface and oxide trap densities. These calculations show that measurable QEH may exist even at the beginning of life, while Heymes et al. (Figure 2) showed that UV-induced surface charge can increase QE by a factor of up to 1.6 at 285nm.[3] The implications for NASA missions can be appreciated in light of the consequences of EUV-induced radiation damage on the calibration of SOHO EIT CCDs.[1]

Radiation-induced charge injection and structural damage to surface dielectrics on silicon have important consequences for field-effect passivation of silicon detectors, which includes surface charging methods such as flash gates, chemisorption, and charged dielectrics (e.g., Al$_2$O$_3$, SiN$_x$, and high-$\kappa$ dielectrics, which are used in solar cells and commercial CMOS imaging detectors). In 2007, MIT Lincoln Labs published a study on CCDs for the EUV Variability Experiment, which showed that ion implantation and MBE-grown silicon are far more radiation-hard than chemisorption passivation.[17]

The QE of a delta-doped detector is plotted in Figure 7 in comparison with calculated QE for a delta-doped detector with an interface trap density of $10^{14}$ cm$^{-2}$ and oxide trap densities varying from $10^{13}$ to $10^{14}$ cm$^{-2}$. Despite the fact that these levels of surface are two orders of magnitude larger than those calculate for the Heymes et al. CCD, the model shows that the QE of a delta-doped detectors is remarkably stable. The tolerance of delta-doped detectors to such extreme levels of radiation damage is explained in Figure 8, which shows that the surface depletion layer is effectively pinned at the position of the first delta layer, independent of variability of interface and oxide trapped charge. Whereas the model predicts residual losses associated with absorption in MBE silicon, data in Figure 7 show that delta-doped detectors respond with nearly 100% internal QE. This discrepancy is attributed to quantum transport in the delta-doped surface, which is not included in the model.[18] In 2012 and 2013, Alacron and Applied Materials verified near-100% QE and radiation-hardness of delta-doped CMOS detectors in months-long accelerated lifetime tests using pulsed excimer lasers at 193nm and 263nm.[18]

Using molecular beam epitaxy (MBE), JPL has developed multilayer 2D-doping to increase surface conductivity (important for high-speed CMOS imaging detectors), compensate defects at the MBE-detector interface, and enhance the stability of delta-doped detectors in high radiation environments. The atomic scale control required to realize these capabilities is the exclusive domain of molecular beam epitaxy and atomic layer deposition technologies developed at JPL's Microdevices Laboratory.[19]

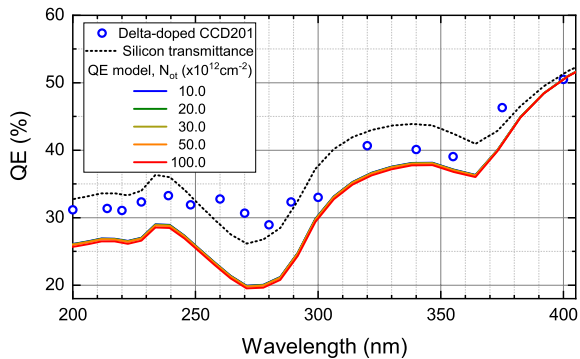 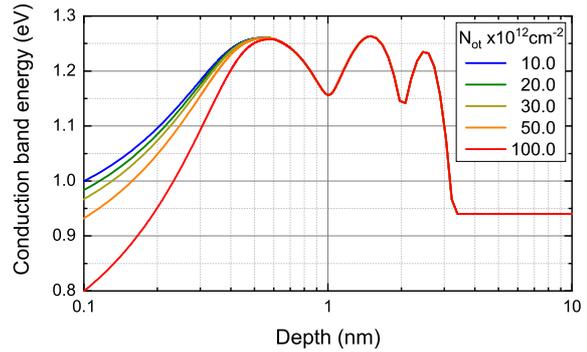

**Figure 7:** The calculated QE of a delta-doped detector is stable against interface trap densities up to $10^{14}cm^{-2}$, in agreement with experiment. QE data for a delta-doped CCD201 are plotted for comparison.[6] The discrepancy between the QE model and data is attributed to quantum transport, which is not included in the model.

**Figure 8:** The conduction band edge of a delta-doped detector is shown here for oxide charge densities in the range of $10^{13}$ - $10^{14}cm^{-2}$. The 2D-dopant profile achieve by MBE growth effectively pins the surface depletion layer at the location of the first delta layer. On these length scales, quantum transport dominates surface interactions.

## Acknowledgements

This work would not have been possible without the exceptional work of scientists in JPL's Advanced UV detectors team: Shouleh Nikzad, April Jewell, John Hennessy, Todd Jones, Gillian Kyne, Chaz Shapiro, and Nathan Bush. I would like to thank David Morris for providing Teledyne e2v's ion implantation profile and for many valuable discussions. The research described in this paper was carried out at the Jet Propulsion Laboratory, California Institute of Technology, under a contract with the National Aeronautics and Space Administration.

## References

[1] Clette *et al.* 2002, "The Radiometric Calibration of the Extreme Ultraviolet Imaging Telescope," *The Radiometric Calibration of SOHO (ESA SR-002)*, Edited by A. Pauluhn, M.C.E. Huber and R. von Steiger.
[2] Jerram *et al.* 2010, "Back-thinned CMOS sensor optimization," *Proc. SPIE* 7598, 759813
[3] Heymes *et al.*, 2020, "Comparison of back-thinned detector ultraviolet quantum efficiency for two commercially available passivation treatments," *IEEE Transactions on Nuclear Science*, 67(8), 1962-1967
[4] Kyne *et al.* 2020, "Delta-doped electron-multiplying CCDs for FIREBall-2," *JATIS* 6(1), 011007
[5] Jewell *et al.* 2018, "Ultraviolet detectors for astrophysics missions: a case study with the star-planet activity research cubesat (SPARC)," *Proc. SPIE* 10709, 107090C
[6] Hoenk *et al.* 2022, "2D-doped silicon detectors for UV/optical/NIR and X-ray astronomy," *SPIE* 12191
[7] Swirhun *et al.* 1986, "Measurement of electron lifetime, electron mobility, and band-gap narrowing in heavily doped p-type silicon," *1986 International Electron Devices Meeting*, Los Angeles, CA, USA, 1986
Swirhun *et al.* 1987, "Temperature dependence of minority carrier electron mobility and bandgap narrowing in p+ silicon," *IEEE Trans. on Electron Dev.* ED-34(11): 2385.
[8] Grove 1967, *Physics and Technology of Semiconductor Devices,* John Wiley and Sons, New York;
[9] McIntosh *et al.* 2014, "On effective surface recombination parameters," *J. Appl. Phys.* 116, 014503
[10] Ragnarrson and Lundgren, 2000, "Electrical characterization of centers in (100) Si-SiO$_2$ structures: The influence of surface potential on passivation during post metallization anneal," *J. Appl. Phys.* **88**, 938
[11] Janesick *et al.*, 1989, "Charge-coupled device pinning technologies," *Proc. SPIE* 1071, 153-169.
[12] Deiries *et al.*, "Precision UV QE measurements of Optical Detectors with a special calibrated work bench," *Scientific Detector Workshop*, Florence, Italy, Oct 7–11, 2013.
[13] Baggett *et al.*, 2010, "WFC3 detectors: On-orbit performance," *Proc. SPIE* 7731, 773138.
[14] Snow et al., 1967, "Effects of ionizing radiation on oxidized silicon surfaces and planar devices," *Proc. IEEE*, 55(7): 1168-1185.
[15] Nissan-Cohen *et al.*, 1986, "Trap generation and occupation dynamics in SiO$_2$ under charge injection stress," *J. Appl. Phys.* **60**(6): 2024-2035.
[16] Afanas'ev *et al.*, "Degradation of the thermal oxide of the Si/SiO$_2$/Al system due to vacuum ultraviolet radiation," *J. Appl. Phys.* **78**(11): 6481-6490, 1995.
[17] Westhoff et al. 2007, "Radiation-hard, charge-coupled devices for the extreme ultraviolet variability experiment," *Proc. SPIE* 6686
[18] Hoenk *et al.* 2015, "Superlattice-doped imaging detectors: progress and prospects," *SPIE Proc.* 9154, High Energy, Optical, and Infrared Detectors for Astronomy VI, 915413, Montreal, Canada (2014).
[19] Nikzad *et al.* 2017, "High Efficiency UV/Optical/NIR Detectors for Large Aperture Telescopes and UV Explorer Missions: Development of and Field Observations with Delta-doped Arrays," *JATIS* 3(3), 03600